\documentclass[letterpaper,english,aps,prl,floatfix,showpacs,twocolumn,amsfonts,amssymb,superscriptaddress,longbibliography]{revtex4-2}

\usepackage{balance}
\usepackage[T1]{fontenc}
\usepackage[utf8]{inputenc}
\usepackage[english]{babel}
\usepackage{graphics}
\usepackage{amssymb}
\usepackage{amsmath, esint}
\usepackage{overpic}
\usepackage[toc]{appendix}
\usepackage{bbold}
\usepackage{comment}

\usepackage{natbib}
\usepackage{xcolor}
\usepackage{soul}
\usepackage{commath}
\usepackage{tikz}
\usepackage{physics}

\usepackage{hyperref}

\newcommand{\be}{\begin{equation}}
\newcommand{\ee}{\end{equation}}
\newcommand{\bspl}{\begin{split}}
\newcommand{\espl}{\end{split}}
\newcommand{\bea}{\begin{eqnarray}}
\newcommand{\eea}{\end{eqnarray}}

\def\nn{\nonumber}
\def\lb{\label}

\newcount\bozza \bozza=0
\ifnum\bozza=1
\newdimen\shift \shift=-2truecm
\def\lb#1{%
{\label{#1}\rlap{\kern\shift{$\scriptstyle#1$}}}}
\else\def\lb#1{\label{#1}} \fi

\definecolor{darkred}{rgb}{0.8, 0.0, 0.0}
\definecolor{darkpowderblue}{rgb}{0.0, 0.2, 0.6}

\usetikzlibrary{shapes.arrows, fadings}
\tikzfading[name=fadeleft,
  left color=transparent!80, right color=transparent!0]

\begin{document}
\title{
 Phonon-Polaritons in Non-Centrosymmetric Systems: Theory of Terahertz Pump- Optical Probe Spectroscopy
}
\author{Niccolò Sellati}
\email{niccolo.sellati@uniroma1.it}
\affiliation{Department of Physics, ``Sapienza'' University of Rome, P.le A.\ Moro 5, 00185 Rome, Italy}
\author{Jacopo Fiore}
\affiliation{Department of Physics, ``Sapienza'' University of Rome, P.le A.\ Moro 5, 00185 Rome, Italy}
\author{Stefano Paolo Villani}
\affiliation{Department of Physics, ``Sapienza'' University of Rome, P.le A.\ Moro 5, 00185 Rome, Italy}
\author{Lara Benfatto}
\affiliation{Department of Physics, ``Sapienza'' University of Rome, P.le A.\ Moro 5, 00185 Rome, Italy}
\author{Mattia Udina}
\email{mattia.udina@uniroma1.it}
\affiliation{Department of Physics, ``Sapienza'' University of Rome, P.le A.\ Moro 5, 00185 Rome, Italy}

\begin{abstract}
Hybrid lattice-light modes, known as phonon-polaritons, represent the backbone of advanced protocols based on THz pumping of infrared modes.  Here we provide a theoretical framework able to capture the different roles played by phonon-polaritons in experimental protocols based either on Raman-like pump and probe schemes, typical of four-wave-mixing processes, or on THz pump-visible probe three-wave mixing protocols. By using a many-body description of the nonlinear optical kernel, along with a perturbative solution of nonlinear Maxwell's equations, we  highlight the advantages of exploiting broadband THz pumps to enlarge the phase space of the phonon-polariton dispersion accessible in a single experiment. Besides providing a quantitative description of existing and future experiments, our results offer a general framework for the theoretical modeling of the hybridization between light and lattice degrees of freedom in time-resolved experiments. 
\end{abstract}
\date{\today}

\maketitle

%
\emph{Introduction.---}
The interaction between infrared-active lattice vibrations and propagating electromagnetic (e.m.)\ fields leads to the emergence of hybrid light-phonon modes in solid state compounds, named phonon-polaritons (PhPs) \cite{basov-review-polaritons,caldwell_nanoph15,lagendijk_prb92,nelson_review,nelson_science03,nelson_review07,kojima_japplphys02}. 
In non-centrosymmetric crystals IR-active phonons can also be Raman-active, and the PhPs can be investigated by means of coherent nonlinear spectroscopy as in four-wave mixing (FWM) or three-wave mixing (TWM) protocols. In the former both the excitation and detection mechanisms rely on a Raman-like coupling of the phonon mode to light, as is the case for Impulsive Stimulated Raman Scattering (ISRS)
\cite{auston_prl85,nelson_optica92,lagendijk_prb92,kurz_prl92,etchepare_prb95,etchepare_jpcm00,johnson_prb18,nelson_jcp02,hebling_prb02,merlin_prb03, takeda_apl15}, where the PhP dispersion is measured by tuning the frequency of a visible pulse and taking advantage of momentum conservation of phase-matched processes. A somehow analogous mechanism underlies older measurements based on nearly-forward spontaneous Raman scattering \cite{henry_prl65,kortus_prb13,loudon_revmodphys72}, where the phase-matching condition was tuned by changing the angle between the incident and scattered light. 
Thanks to the huge technical improvements in the generation of intense THz light pulses \cite{giorgianni_narrow,pezeril_optica18} the detection of PhPs is nowadays possible also via THz pump-optical probe TWM processes. In this case the IR excitation of the phonon can be achieved with a THz pulse, and the PhP can be detected as reflectivity or transmission changes of a time-delayed optical probe via a Raman-like interaction \cite{johnson_apl15,kojima_photonics18,johnson_apl17,johnson_prl19,johnson_jap19,blake_prl22,yamaguchi_scirep16,giorgianni_narrow}. A similar TWM mechanism has been exploited in the past in Raman experiments, by simultaneously irradiating the sample with a continuous-wave IR laser \cite{henry_prl66,hwang_ssc73,hwang_prb74,claus_pssa81,keilmann_prb86,keilmann_prl03}. In this configuration, the main observation is that PhPs with a finite momentum mismatch can be excited and detected, even though with a still unexplained marked modulation of the detection signal. 
So far, a complete theoretical characterization of the role played by the hybrid character of the PhP, the phase mismatch and the propagation effects in the different multiwave protocols has not been provided yet, hindering the potential applications of PhPs in phonon-driven phase transitions and in their interaction with additional collective modes \cite{nelson_optica92,weiner_prb95,kurz_prb94,kurz_RevModPhys98,cavalleri_xray_nature06,cavalleri_prl17,cavalleri_natphys22,cavalleri_cm24,gedik_natcomm24}.
%
%

In this work we develop a general theoretical framework to describe the detection of PhPs in non-centrosymmetric cubic systems through FWM and TWM protocols. We combine a many-body derivation of the nonlinear current involved in multiwave mixing processes with a perturbative solution of Maxwell's equations in the presence of a nonlinear source current, in order to correctly describe propagation effects related to the phase-matching conditions. We show that in the FWM case one directly accesses the phonon component of the PhP, explaining both stimulated Raman, such as ISRS \cite{auston_prl85,nelson_optica92,lagendijk_prb92,kurz_prl92,etchepare_prb95,etchepare_jpcm00,nelson_jcp02,hebling_prb02,merlin_prb03, takeda_apl15,johnson_prb18}, as well as spontaneous Raman \cite{henry_prl65,kortus_prb13,loudon_revmodphys72} experiments. The formalism is then applied to TWM protocols based on THz pump-optical probe experiments with a narrowband (multi-cycle) or broadband (single-cycle) pump. We demonstrate that in this case the PhP dispersion enters mainly via the propagation effects of the nonlinear signal generated by the THz pump. For narrowband pulses, that closely resemble Raman experiments stimulated with continuous-wave IR radiation \cite{henry_prl66,hwang_ssc73,hwang_prb74,claus_pssa81,keilmann_prb86,keilmann_prl03}, we explain how the signal modulation depends on the generation of PhPs with a finite momentum mismatch. For broadband pulses we show how transmission measurements can access the polariton dispersion by tuning the spectral features of the pump and probe pulses, making TWM a preferential knob to investigate the light-matter hybridization process. 

\emph{Many-body approach.---} As a starting point we model the IR-active phonon mode as an harmonic oscillator of frequency $\omega_\text{TO}$. Given the phonon displacement $\textbf Q(i\Omega_m,\textbf k)$ and phonon momentum $\textbf P(i\Omega_m,\textbf k)$, with $\Omega_m=2\pi m/\beta$ the bosonic Matsubara frequencies and $\beta$ the inverse temperature, one can write the Gaussian action in the path integral formulation \cite{nagaosa} as $S[\textbf Q,\textbf P]=\sum\limits_{i\Omega_m,\textbf k}\big[H(i\Omega_m,\textbf k)+\Omega_m \textbf P(i\Omega_m,\textbf k)\cdot \textbf Q(-i\Omega_m,-\textbf k)\big]$,
%
%
where $H(i\Omega_m,\textbf k)=\omega_{\text{TO}}^2|\textbf Q(i\Omega_m,\textbf k)|^2/2+|\textbf P(i\Omega_m,\textbf k)|^2/2$ gives the phonon Hamiltonian.
By integrating out $\textbf P$, one is left with $S_0[\textbf Q]=\sum_{i\Omega_m,\textbf k}[D_0^{-1}|\textbf Q(i\Omega_m,\textbf k)|^2]$, where $D_0(i\Omega_m)=2/(\omega_\text{TO}^2+\Omega_m^2)$ is the bare phonon propagator. To model multiwave detection processes, we must include all possible couplings between the phonon and the e.m.\ field. Since the mode is IR-active, the system admits a linear coupling to the gauge field $\textbf A$ that can be included via the minimal coupling substitution \cite{mahan} $\textbf P(i\Omega_m,\textbf k)\to \textbf P(i\Omega_m,\textbf k)-\text Z\textbf{A}(i\Omega_m,\textbf k)/c$ in $H(i\Omega_m,\textbf k)$, where $c$ is the speed of light and $\text Z$ is a rank-2 tensor that can be connected to the Born effective charge. In cubic crystals $\text Z$ is scalar, such that longitudinal and transverse components are decoupled. Integrating $\textbf P$ one is left with $S_\text{IR}[\textbf Q,\textbf A]=\sum_{i\Omega_m,\textbf k}[\Omega_m(\text Z\textbf A/c)\cdot\textbf Q]$. From now on we only focus on the coupling of the transverse components of the phonon $\textbf Q_T$ to the e.m.\ field. Whenever the mode is also Raman-active one should add an additional quadratic (Raman-like) coupling term
$S_R[\textbf Q_T,\textbf A]=\sum_{i\Omega_m,\textbf k}\sum_{i\Omega_n,\textbf k'}[\textbf Q_T\mathcal R\textbf A\textbf A]$ \cite{udina_prb19}, where $\mathcal R$ is a rank-3 tensor that can be connected to the phonon Raman tensor. Finally, one should include the quantum action $S_\text{e.m.}[\textbf A]$ describing the fluctuations of the e.m.\ field. In the presence of external perturbations some spectral components of $\textbf A$ are connected to the pump or probe fields, and one is interested in computing a nonlinear current $\textbf J^{(\text n)}$ (with $\text n=2$ for TWM and $\text n=3$ for FWM) with respect to them by integration of any other fluctuating components.

\emph{Four-wave mixing interaction.---} To clarify the approach, let us start from FWM processes. In this case the pump and the probe are represented by a visible field $\textbf A_R$. The THz field, linearly coupled to the phonon, can be integrated out as its spectral components are well separated from those of $\textbf A_R$. Such procedure, detailed in \cite{suppl}, has the primary effect of dressing the bare phonon with the THz e.m.\ fields, leading to a PhP mode coupled to the external perturbation $\textbf A_R$, $S[\textbf Q_T,\textbf A_R]=S_G[\textbf Q_T]+S_{R}[\textbf Q_T,\textbf A_R]$, where $S_G[\textbf Q_T]=\sum_{i\Omega_m,\textbf k}[D^{-1}_\text{Q}\big|\textbf Q_T(i\Omega_m,\textbf k)\big|^2]$ and 
%
%
\begin{align}\lb{dressprop}
    D_\text{Q}(i\Omega_m,\textbf k)=\frac{2}{\omega_\text{TO}^2+\Omega_m^2+\frac{\Omega_P^2\Omega_m^2}{\Omega_m^2+\frac{c^2}{\varepsilon_\infty}|\textbf k|^2}}
\end{align}
is the propagator of the phonon dressed by the interaction with light, see Fig.\ \ref{fig1}(a,b). In this expression $\varepsilon_\infty$ is the high-frequency dielectric constant and $\Omega_P^2$ is the ionic plasma frequency proportional to the matrix elements of $\text Z^2$, entering the definition of the longitudinal-optical phonon frequency as $\omega_{\text{LO}}^2=\omega_{\text{TO}}^2+\Omega_P^2$. Eq.\ \eqref{dressprop} can equivalently be written as $D_\text{Q}=D_0(\varepsilon_\infty\Omega_m^2+c^2|\textbf k|^2)/(\varepsilon(i\Omega_m)\Omega_m^2+c^2|\textbf k|^2)$ where the dielectric function reads $\varepsilon(i\Omega_m)=\varepsilon_\infty\big(\Omega_m^2+\omega_{\text{LO}}^2\big)/\big(\Omega_m^2+\omega_{\text{TO}}^2\big)$. 
Once the analytical continuation to real frequencies $i\Omega_m\to\omega+i0^+$ is performed, the poles of the propagator identify the dispersion of the phonon mode. One then recovers the two polaritonic branches, given by the wave equation $\varepsilon(\omega)\omega^2=c^2|\textbf k|^2$ \cite{mahan}, weighted by the factor $\varepsilon_\infty\omega^2-c^2|\textbf k|^2$ that accounts for the phonon nature of the polariton. This weight affects the intensity of the phonon spectral function $G_\text{Q}(\omega,\textbf k)=-\text{Im}[D_\text{Q}(i\Omega_m\to\omega+i\gamma,\textbf k)]/\pi$ shown in Fig.\ \ref{fig1}(a), where we introduced a  finite imaginary part $i0^+\to i\gamma$ to account for the broadening of the dispersion due to disorder or correlation effects. 
\begin{figure}[t!]
    \centering
    \includegraphics[width=0.48\textwidth,keepaspectratio]{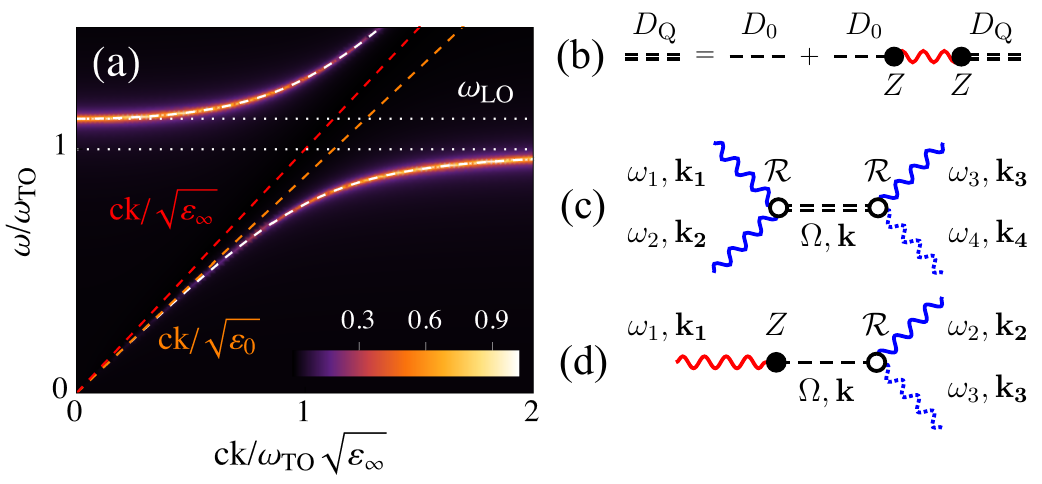}
    \caption{(a) Absolute value of the PhP spectral function, obtained from Eq.\ \eqref{dressprop}, normalized to its maximum value. Horizontal white dotted lines highlight the $\omega_{\text{TO}}$ and $\omega_{\text{LO}}$ frequencies (here $\omega_{\text{LO}}=1.125\omega_{\text{TO}}$). White dashed curves are the polaritonic dispersion branches. Here we set $\gamma=0.01\omega_\text{TO}$. (b) Dyson equation for $D_\text{Q}$ (double dashed line), obtained by integrating out the e.m.\ field (red line) to dress the bare phonon propagator $D_0$ (single dashed line). (c) FWM interaction between three optical photons $\textbf A_R$ (blue lines), mediated by the PhP. (d) TWM interaction between the THz field $\textbf A_p$ (red line) and the optical field $\textbf A_R$ (blue line), mediated by the bare phonon. The nonlinear current $\textbf J^{(\text n)}$ related to the two processes is obtained as the functional derivative with respect to the outgoing optical field (dashed blue line).}
    \lb{fig1}
\end{figure}\\
By integrating the phonon out of the action $S[\textbf Q_T,\textbf A_R]$, one then finds the effective-action $S_\text{eff}[\textbf A_R]$ for the external e.m.\ field, that is of fourth order in $\textbf A_R$. We consider the case in which the polariton at frequency $\Omega=\omega_1+\omega_2$ and momentum $\textbf k=\textbf k_1+\textbf k_2$ is first excited by optical rectification of two incoming optical photons $\textbf A_R(\omega_1,\textbf k_1)$ and $\textbf{A}_R(\omega_2,\textbf k_2)$, with $\text k_i=n(\omega_i)\omega_i/c$ and $n(\omega)$ the refractive index, and subsequently scatters with a third photon $\textbf A_R(\omega_3,\textbf k_3)$, see Fig.\ \ref{fig1}(c). 
%
The functional derivative of the effective action with respect to the scattered optical field $\textbf A_R(\omega_4,\textbf k_4)$ gives the outgoing nonlinear current, i.e.\ $\textbf J^{(3)}(\omega_4,\textbf k_4)=\int \textbf A_R(\omega_1,\textbf k_1)\textbf A_R(\omega_2,\textbf k_2)\text K^{(3)}(\omega_1+\omega_2,\textbf k_1+\textbf k_2)\textbf A_R(\omega_3,\textbf k_3)\delta(\omega_1+\omega_2+\omega_3-\omega_4)d\omega_1d\omega_2d\omega_3$, where 
\begin{align}\lb{ker}
\text K^{(3)}(\Omega,\textbf k)\propto \mathcal{R}^2 D_\text{Q}(\Omega,\textbf k)
\end{align}
is the third-order nonlinear kernel. Notice that since $\text K^{(3)}$ is proportional to the PhP propagator Eq.\ \eqref{dressprop}, the current is non-negligible only when the hybrid mode has sizeable phonon character at frequency $\Omega=\omega_1+\omega_2$. 

\emph{Three-wave mixing interaction.---}
In the TWM case the coherent excitation of the IR-active phonon is accomplished by an external THz field $\textbf A_p$ resonant with the lattice mode. In this case the THz field cannot be integrated out of the action, since one is interested in a nonlinear current $\textbf J^{(2)}$ that scales linearly with $\textbf A_p$.
By taking into account a Raman-like detection with a visible probe field $\textbf A_R$, the full TWM light-phonon action reads $S[\textbf Q_T,\textbf A_R,\textbf A_p]=S_0[\textbf Q_T]+S_R[\textbf Q_T,\textbf A_R]+S_\text{IR}[\textbf Q_T,\textbf A_p]$. By integrating out the transverse bare phonon one finds the effective-action $S_\text{eff}[\textbf A_R, \textbf A_p]$ for the e.m.\ fields, from which we retain only terms at first order in the THz pump $\textbf A_p(\omega_1,\textbf k_1)$ and second order in the optical field $\textbf A_R(\omega_2,\textbf k_2)\textbf A_R(\omega_3,\textbf k_3)$, see Fig.\ \ref{fig1}(d). The nonlinear current reads $\textbf J^{(2)}(\omega_3,\textbf k_3)=\int \textbf A_p(\omega_1,\textbf k_1)\text K^{(2)}(\omega_1)\textbf A_R(\omega_2,\textbf k_2)\delta(\omega_1+\omega_2-\omega_3)d\omega_1d\omega_2$, where the second-order nonlinear kernel
\begin{align}\lb{kerPP}
    \text K^{(2)}(\Omega)\propto\Omega(\text Z\mathcal R)D_0(\Omega)
\end{align}
scales as a rank-3 tensor, coming from the interplay between the tensors $\text Z$ and $\mathcal R$, mediated by the non-centrosymmetric bare phonon mode. The tensorial structure of the kernel is closely related to the symmetry properties of the sample \cite{hwang_prb74,merlin_prb03,kortus_prb13}, and it determines the selection rules for the corresponding nonlinear response. Indeed, as we detail in \cite{suppl}, the signal can be strongly suppressed, and even disappear, in specific polarization geometries according to the structure of the crystal \cite{yamaguchi_scirep16}. In contrast to the FWM case, the nonlinear kernel in Eq.\ \eqref{kerPP} is proportional to the \textit{bare} phonon propagator. Nonetheless, as we shall discuss below, the polariton affects the response via the propagation effects of the THz field $\textbf A_p$, that directly controls the nonlinear current $\textbf J^{(2)}$. 

\emph{Propagation effects.---}
The nth-order nonlinear current $\textbf{J}^\text{(n)}(\omega,z)$ acts as a source term in Maxwell's equation for the gauge field, \ $\partial_z^2\textbf A(\omega,z)+n^2(\omega)\omega^2\textbf A(\omega,z)/c^2=-4\pi\textbf J^\text{(n)}(\omega,z)/c$. By treating the nonlinear current as a perturbation, we derive the leading-order perturbative solution $\textbf A^{[1]}(\omega,z)$ in a confined region $0<z<d$, where $d$ is the sample thickness. The procedure is detailed in \cite{suppl}. 
The solution explicitly accounts for a possible finite momentum mismatch $\Delta \text k$ between the interacting e.m.\ fields.
When computing e.g.\ the transmitted field $\textbf A^{[1]}(\omega,d^+)$ in the TWM case, the phase-matching condition $\Delta \text k = 0$ requires $n(\omega_1)\omega_1/c=n(\omega)\omega/c- n(\omega_2)\omega_2/c$. Here $\omega_1$ denotes the frequency of the incoming THz field while $\omega_2$ and $\omega$ denote the frequencies of the incoming and outgoing visible light pulses respectively, with $\omega=\omega_1+\omega_2$. Considering an approximately constant optical refractive index $n(\omega)\simeq n(\omega_2)\equiv n_\text{eV}$, the phase-matching condition is $n(\omega_1)\omega_1/c=n_\text{eV}\omega_1/c$, which admits a simple graphical representation in the $(\omega_1,\text k_1)$-plane as the intersection between the PhP dispersion and the curve $\omega_1=c\text k_1/n_\text{eV}$ \cite{henry_prl65}, which typically occurs at momenta in which the PhP is of hybrid character. If one is interested in the reflected field $\textbf A^{[1]}(\omega,0^-)$, instead, one should replace $\omega_2\to-\omega_2$ in the phase-matching condition and the two curves intersect at high momenta where the hybrid PhP character is lost. 

\emph{Pump-probe experiments.---}
In pump-probe protocols variations in the probe field are collected as function of the pump-probe time delay $t_{pp}$,
at fixed observation time. As detailed in \cite{suppl}, the Fourier transform of the measured signal in transmission $\textbf{A}_{\text{tr}}(\omega_{pp})$ or reflection $\textbf{A}_{\text{ref}}(\omega_{pp})$ geometry can be directly related to $\textbf A^{[1]}(\omega,d^+)$ and $\textbf A^{[1]}(\omega,0^-)$ respectively. 
\begin{figure}[t!]
    \centering
    \includegraphics[width=0.48\textwidth,keepaspectratio]{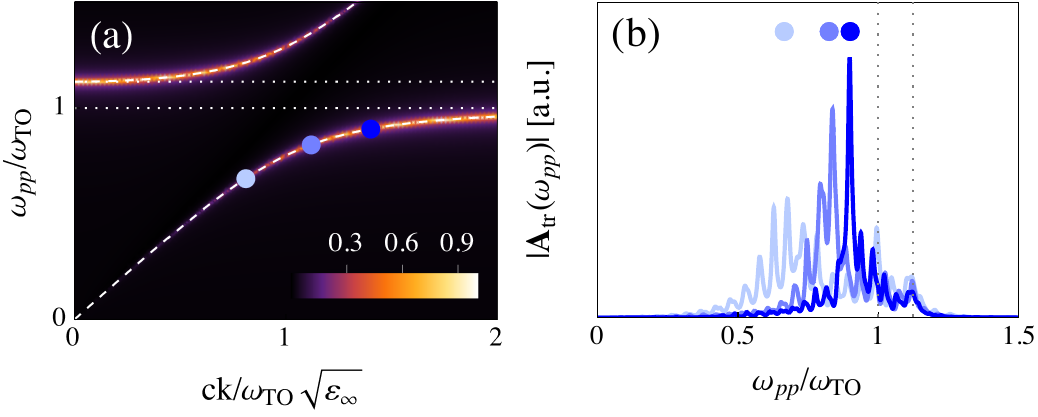}
    \caption{Optical pump-optical probe experiments in transmission geometry with collinear pulses. (a) Phase-matched points for different values of $n_\text{eV}/\sqrt{\varepsilon_\infty}$, changed between $1.20$ (light blue), $1.35$ (blue) and $1.54$ (dark blue). (b) Corresponding transmitted intensity $|\textbf A_\text{tr}(\omega_{pp})|$. In both panels, $\omega_\text{LO}=1.125\omega_\text{TO}$, $\gamma=0.01\omega_\text{TO}$ and $d=50c/\omega_\text{TO}\sqrt{\varepsilon_\infty}$ (order of $100\text{ }\mu \text{m}$ in typical samples). Dotted lines highlight $\omega_\text{TO}$ and $\omega_\text{LO}$.}
    \lb{fig2}
\end{figure}\\
We first consider optical pump-optical probe FWM experiments (ISRS). In this case, the nonlinear current is a convolution of the kernel $\text K^{(3)}$ in Eq.\ \eqref{ker} with the optical e.m.\ field $\textbf{A}_R$ inside the sample. At visible frequencies $\omega\gg\omega_\text{TO}$ the PhP does not quantitatively affect the refractive index $n(\omega)$, whose dispersion is essentially controlled by electronic processes. Because of this, the resulting nonlinear signal can be understood as a direct measurement of the phonon character of the PhP encoded in the dressed phonon propagator Eq.\ \eqref{dressprop}, with momentum dictated by the phase-matching condition $\Delta \text k=0$ that allows one to reconstruct the dispersion \cite{auston_prl85,nelson_optica92,lagendijk_prb92,kurz_prl92,etchepare_prb95,etchepare_jpcm00,nelson_jcp02,hebling_prb02,merlin_prb03, takeda_apl15,johnson_prb18}. 
Fig.\ \ref{fig2} shows a qualitative result for $|\textbf A_\text{tr}(\omega_{pp})|$ in optical pump-probe experiments. 
The phase-matched frequency is tuned by changing the central frequency of the pulses, and thus $n_\text{eV}$. Notice that the intensity of the main peak grows approaching $\omega_\text{TO}$, as the phonon character of the PhP progressively increases. The satellite peaks are associated with screening effects at the interface and Fabry-Perot internal reflections.
%
%
%

In THz pump-optical probe TWM experiments the presence of the PhP makes the refractive index for the THz pump pulse $\textbf A_p$ highly dispersive. Because of this, 
even though the polariton does not explicitly appear in the $\text K^{(2)}$ kernel Eq.\ \eqref{kerPP}, it still modulates the propagation of the THz field inside the material and, consequently, of the generated nonlinear signal. We model the incident THz pump field as a Gaussian spectrum centered at $\omega_p$ with pulse duration $\tau$. The polarization of the pump and probe pulses are fixed
to maximize the signal according to the symmetry properties of the kernel Eq.\ \eqref{kerPP}, while the probe frequency is tuned to adjust the phase-matching condition $\Delta\text{k}=0$ and move along the dispersion branches. As discussed above, to access the PhP dispersion in the region where the light-phonon mixing is strong, 
experiments in transmission geometry are preferable as the phase-matching condition is usually realized at small momenta $\text k\sim \sqrt{\varepsilon_\infty}\omega_\text{TO}/c$. In this case the signal in frequency space reads
\begin{align}\lb{atr}
    \textbf A_\text{tr}&(\omega_{pp})=\sum_{\sigma_1,\sigma_2}^{\pm 1}\textbf A_{\sigma_1}(\omega_{pp})\text K^{(2)}(\omega_{pp})\int d\omega \textbf A_{\sigma_2}(\omega-\omega_{pp})\nn\\
    &\times\frac{f(\omega)t(\omega)}{\omega}e^{i\frac{(n(\omega)+1)\omega d}{c}}\bigg[\sum_{\alpha}^{\pm 1}r_\alpha(\omega)\frac{1-e^{i\Delta \text k_\alpha d}}{\Delta \text k_\alpha}\bigg],
\end{align}
where the coefficients $t(\omega)=2/(n(\omega)+1)$, $r_{\pm}(\omega)=(n(\omega)\pm1)/(n(\omega)+1)$ and the Fabry-Perot factor $f(\omega)=1/(1-r_-^2(\omega)e^{2i n(\omega)\omega d/c})$ account for the propagation of the e.m.\ fields at the interfaces of the sample, $\textbf A_{+1}(\omega)=\textbf A_t(\omega)=\textbf A^\text{ext}(\omega)t(\omega)f(\omega)$ and $\textbf A_{-1}(\omega)=\textbf A_r(\omega)=\textbf A^\text{ext}(\omega)t(\omega)r_-(\omega)f(\omega)e^{2in(\omega)\omega d/c}$, with $\textbf A^\text{ext}(\omega)=\textbf A_p^\text{ext}(\omega)+\textbf A_R^\text{ext}(\omega)$ the external perturbation. The term in the square bracket is the phase-matching factor, where the momentum mismatch is $\Delta \text k_\alpha=\sum_i \sigma_i\text k_i-\alpha n(\omega)\omega/c$, with $\text k_1=n(\omega_{pp})\omega_{pp}/c$ and $\text k_2=n(\omega-\omega_{pp})(\omega-\omega_{pp})/c$.
\begin{figure}[t!]
    \centering
    \includegraphics[width=0.48\textwidth,keepaspectratio]{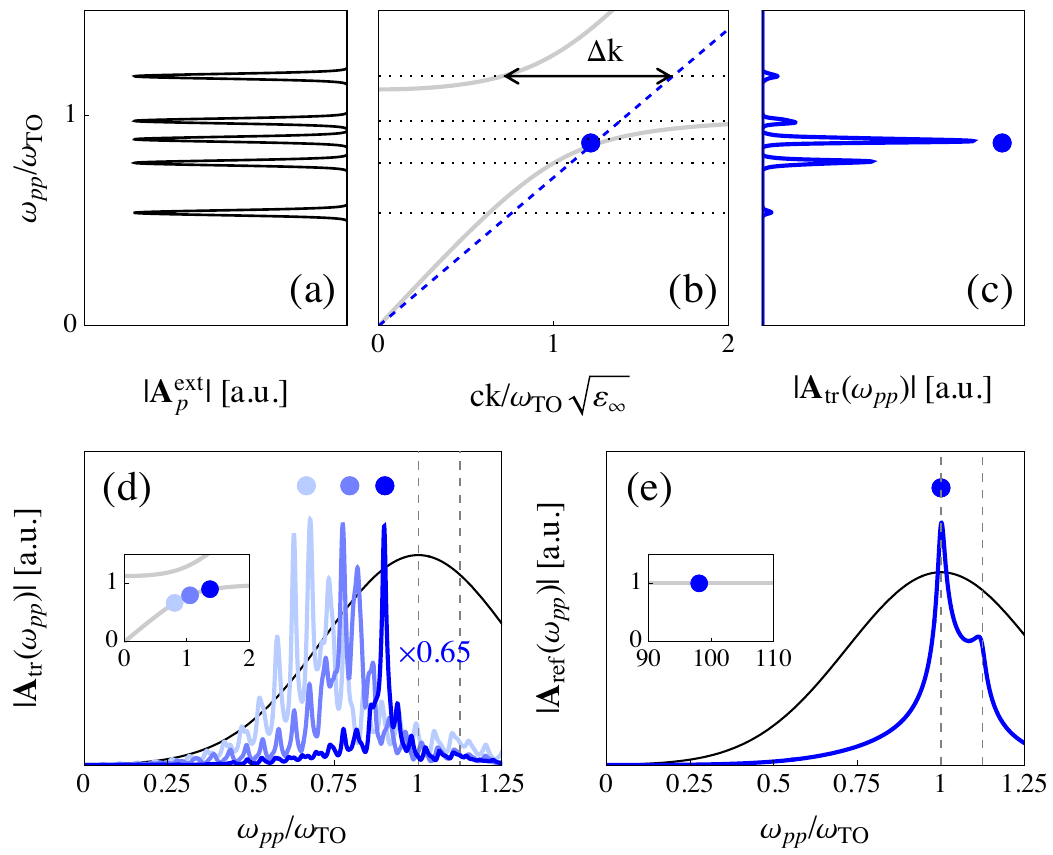}
    \caption{THz pump-optical probe experiments with collinear pulses. (a) Spectral content of narrowband pump pulses with different central frequencies $\omega_p$. (b) Phase-matched point in transmission geometry identified by the intersection between the PhP dispersion (gray) and $c\text k/n_\text{eV}$ (blue dashed). Dotted horizontal lines are guides to the eye at $\omega_p$. Black arrow represents the momentum mismatch $\Delta\text k$ for one of the pumps. (c) Transmitted signal $|\textbf A_\text{tr}(\omega_{pp})|$ for the different pump pulses shown in panel (a). The blue dot marks the phase-matched frequency. (d) Transmitted signal $|\textbf A_\text{tr}(\omega_{pp})|$ induced by a broadband pump pulse (black). $n_\text{eV}/\sqrt{\varepsilon_\infty}$ is changed between $1.20$ (light blue), $1.32$ (blue) and $1.54$ (dark blue) to move the phase-matched frequency (dots) along the dispersion (inset). Vertical dashed lines highlight $\omega_\text{TO}$ and $\omega_\text{LO}$. (e) Reflected signal $|\textbf A_\text{ref}(\omega_{pp})|$ induced by a broadband pump pulse (black). The phase-matched frequency (dot) is fixed by $n_\text{eV}/\sqrt{\varepsilon_\infty}=1.41$ and falls on the high-momentum region of the dispersion (inset). Vertical dashed lines highlight $\omega_\text{TO}$ and $\omega_\text{LO}$. In all panels, $\omega_\text{LO}=1.125\omega_\text{TO}$ and
    $\gamma=0.01\omega_\text{TO}$. Sample thickness is $d=50c/\omega_\text{TO}\sqrt{\varepsilon_\infty}$ for transmission and $d\to \infty$ for reflection geometries.}
    \lb{fig3}
\end{figure}
In Fig.\ \ref{fig3}(a-c) we show the signal intensity $|\textbf A_\text{tr}(\omega_{pp})|$ induced by narrowband THz pump pulses ($\omega_p\tau\gg1$) with different central frequencies $\omega_p$, while $n_\text{eV}$ is kept fixed.
As expected, the response is maximized when the phase-matching condition is fulfilled. Nonetheless, a sizable response is also observed for $\Delta \text{k} \ne 0$, underlining the main difference between the coherent nonlinear excitation of PhPs and spontaneous Raman scattering, in which only phase-matched polaritons can be observed \cite{henry_prl65,kortus_prb13,loudon_revmodphys72}. 
We also point out that THz pump-optical probe measurements in the presence of a monochromatic ($\omega_p\tau\to+\infty$) THz pump can be mapped into Raman measurements stimulated by coherent IR radiation \cite{henry_prl66,hwang_ssc73,hwang_prb74,claus_pssa81,keilmann_prb86,keilmann_prl03}. In particular, our results are in good agreement with Ref.\ \cite{henry_prl66}, where a sizable response is observed even when the frequency $\omega_p$ explores the upper polariton branch.\\
When broadband THz pump pulses ($\omega_p\tau\ll1$) are applied, one can in principle access the full frequency range in which the polariton has mixed light-matter character within a single measurement. Fig.\ \ref{fig3}(d) shows $|\textbf A_\text{tr}(\omega_{pp})|$ for a fixed pump field and different optical refractive indices. 
Analogously to the ISRS case, the main frequency of the oscillations is the phase-matched one, redshifted with respect to $\omega_{\text{TO}}$. Nonetheless, in this case the relative height of the peaks cannot be directly linked to the phonon character of the PhP.
It should be noted that in thin samples ($d\lesssim c/\omega_\text{TO}\sqrt{\varepsilon_\infty}$) or when the penetration depth of the optical pulses is small ($\delta\lesssim c/\omega_\text{TO}\sqrt{\varepsilon_\infty}$) the phase-matching condition loses significance, see \cite{suppl} for further details.\\
In Fig.\ \ref{fig3}(e) we show the reflected signal $|\textbf A_\text{ref}(\omega_{pp})|$, given by a broadband THz pump. In this case the phase-matching condition is met in the region $\text k\gg \sqrt{\varepsilon_\infty}\omega_\text{TO}/c$ in which the dispersion of the PhP becomes flat and approaches the bare-phonon frequency $\omega_\text{TO}$. As a consequence, the signal is peaked at $\omega_\text{TO}$, while a secondary peak at $\omega_\text{LO}$ is given by the screening of the THz pump at the interface, in excellent agreement with experimental data \cite{yamaguchi_scirep16}.
%
%
%

\emph{Summary and discussion.---} 
In this work we provided a microscopic derivation of the nonlinear signal generated by multiwave mixing processes 
in non-centrosymmetric cubic crystals. We rely on a microscopic derivation of the nonlinear current generated in the sample, based on a many-body effective-action derivation of the optical kernel, and on a perturbative solution of the nonlinear Maxwell's equations, needed to account for propagation effects and phase-matching conditions. We have shown that the hybrid nature of the PhP enters the response in different ways depending on the experimental protocols.
For all-optical FWM interactions we showed that the nonlinear kernel itself maps out the {\em phonon} component of the PhP, see Eq.\ \eqref{ker}, while propagation effects play a minor role. This allows one to directly access the hybrid-mode dispersion by tuning the frequency of the visible probe pulse in ISRS measurements. However, when the phase-matching condition moves towards the low-momenta region the signal looses intensity, due to the suppression of the phonon character of the PhP branch, see Fig.\ \ref{fig2}. In contrast, for THz pump-optical probe TWM protocols the nonlinear kernel itself is proportional to the bare phonon mode, see Eq.\ \eqref{kerPP}, and light-phonon hybridization affects the propagation of the THz pump field, leading again to a pronounced peak at the phase-matched frequency. This has the remarkable advantage that for broadband THz pulses one can push the detection towards the low-momenta region, enlarging the phase space of the PhP dispersion accessible with the same experimental configuration.  
On a general perspective, the present formalism
is flexible to include additional coupling channels between the phonon and other matter degrees of freedom, as e.g.\ anharmonically coupled PhPs in anisotropic crystals \cite{lagendijk_prb92,johnson_apl17,johnson_prb18,blake_prl22,etchepare_prb95,etchepare_jpcm00,johnson_jap19,johnson_prl19}, or magnon-phonon-polaritons \cite{nelson_magn_pol}, as well as to describe higher order IR-phonon excitations
\cite{forst2011, IKE_24}. Finally, since the effective-action description incorporates the phonon and light fields on the same footing, our results can be generalized to infer information on light-matter interaction in optical cavities, given the intriguing possibility to investigate cavity polaritons using pump-probe spectroscopy \cite{delpo20,fassioli_jpcl21}.\\

We acknowledge financial support by EU under project MORE-TEM ERC-SYN (No.\ 951215),  by Sapienza University under projects
Ateneo (RP1221816662A977 and RM123188E357C540) and by ICSC–Centro Nazionale di Ricerca in High Performance
Computing, Big Data and Quantum Computing, funded by European Union – NextGenerationEU. 

\bibliography{bibl.bib} 

\begin{thebibliography}{52}%
\makeatletter
\providecommand \@ifxundefined [1]{%
 \@ifx{#1\undefined}
}%
\providecommand \@ifnum [1]{%
 \ifnum #1\expandafter \@firstoftwo
 \else \expandafter \@secondoftwo
 \fi
}%
\providecommand \@ifx [1]{%
 \ifx #1\expandafter \@firstoftwo
 \else \expandafter \@secondoftwo
 \fi
}%
\providecommand \natexlab [1]{#1}%
\providecommand \enquote  [1]{``#1''}%
\providecommand \bibnamefont  [1]{#1}%
\providecommand \bibfnamefont [1]{#1}%
\providecommand \citenamefont [1]{#1}%
\providecommand \href@noop [0]{\@secondoftwo}%
\providecommand \href [0]{\begingroup \@sanitize@url \@href}%
\providecommand \@href[1]{\@@startlink{#1}\@@href}%
\providecommand \@@href[1]{\endgroup#1\@@endlink}%
\providecommand \@sanitize@url [0]{\catcode `\\12\catcode `\$12\catcode `\&12\catcode `\#12\catcode `\^12\catcode `\_12\catcode `\%12\relax}%
\providecommand \@@startlink[1]{}%
\providecommand \@@endlink[0]{}%
\providecommand \url  [0]{\begingroup\@sanitize@url \@url }%
\providecommand \@url [1]{\endgroup\@href {#1}{\urlprefix }}%
\providecommand \urlprefix  [0]{URL }%
\providecommand \Eprint [0]{\href }%
\providecommand \doibase [0]{http://dx.doi.org/}%
\providecommand \selectlanguage [0]{\@gobble}%
\providecommand \bibinfo  [0]{\@secondoftwo}%
\providecommand \bibfield  [0]{\@secondoftwo}%
\providecommand \translation [1]{[#1]}%
\providecommand \BibitemOpen [0]{}%
\providecommand \bibitemStop [0]{}%
\providecommand \bibitemNoStop [0]{.\EOS\space}%
\providecommand \EOS [0]{\spacefactor3000\relax}%
\providecommand \BibitemShut  [1]{\csname bibitem#1\endcsname}%
\let\auto@bib@innerbib\@empty
\bibitem [{\citenamefont {Basov}\ \emph {et~al.}(2016)\citenamefont {Basov}, \citenamefont {Fogler},\ and\ \citenamefont {de~Abajo}}]{basov-review-polaritons}%
  \BibitemOpen
  \bibfield  {author} {\bibinfo {author} {\bibfnamefont {D.~N.}\ \bibnamefont {Basov}}, \bibinfo {author} {\bibfnamefont {M.~M.}\ \bibnamefont {Fogler}}, \ and\ \bibinfo {author} {\bibfnamefont {F.~J.~G.}\ \bibnamefont {de~Abajo}},\ }\href {\doibase 10.1126/science.aag1992} {\bibfield  {journal} {\bibinfo  {journal} {Science}\ }\textbf {\bibinfo {volume} {354}},\ \bibinfo {pages} {aag1992} (\bibinfo {year} {2016})}\BibitemShut {NoStop}%
\bibitem [{\citenamefont {Caldwell}\ \emph {et~al.}(2015)\citenamefont {Caldwell}, \citenamefont {Lindsay}, \citenamefont {Giannini}, \citenamefont {Vurgaftman}, \citenamefont {Reinecke}, \citenamefont {Maier},\ and\ \citenamefont {Glembocki}}]{caldwell_nanoph15}%
  \BibitemOpen
  \bibfield  {author} {\bibinfo {author} {\bibfnamefont {J.~D.}\ \bibnamefont {Caldwell}}, \bibinfo {author} {\bibfnamefont {L.}~\bibnamefont {Lindsay}}, \bibinfo {author} {\bibfnamefont {V.}~\bibnamefont {Giannini}}, \bibinfo {author} {\bibfnamefont {I.}~\bibnamefont {Vurgaftman}}, \bibinfo {author} {\bibfnamefont {T.~L.}\ \bibnamefont {Reinecke}}, \bibinfo {author} {\bibfnamefont {S.~A.}\ \bibnamefont {Maier}}, \ and\ \bibinfo {author} {\bibfnamefont {O.~J.}\ \bibnamefont {Glembocki}},\ }\href {\doibase doi:10.1515/nanoph-2014-0003} {\bibfield  {journal} {\bibinfo  {journal} {Nanophotonics}\ }\textbf {\bibinfo {volume} {4}},\ \bibinfo {pages} {44} (\bibinfo {year} {2015})}\BibitemShut {NoStop}%
\bibitem [{\citenamefont {Planken}\ \emph {et~al.}(1992)\citenamefont {Planken}, \citenamefont {Noordam}, \citenamefont {Kennis},\ and\ \citenamefont {Lagendijk}}]{lagendijk_prb92}%
  \BibitemOpen
  \bibfield  {author} {\bibinfo {author} {\bibfnamefont {P.~C.~M.}\ \bibnamefont {Planken}}, \bibinfo {author} {\bibfnamefont {L.~D.}\ \bibnamefont {Noordam}}, \bibinfo {author} {\bibfnamefont {J.~T.~M.}\ \bibnamefont {Kennis}}, \ and\ \bibinfo {author} {\bibfnamefont {A.}~\bibnamefont {Lagendijk}},\ }\href {\doibase 10.1103/PhysRevB.45.7106} {\bibfield  {journal} {\bibinfo  {journal} {Phys. Rev. B}\ }\textbf {\bibinfo {volume} {45}},\ \bibinfo {pages} {7106} (\bibinfo {year} {1992})}\BibitemShut {NoStop}%
\bibitem [{\citenamefont {Feurer}\ \emph {et~al.}(2007{\natexlab{a}})\citenamefont {Feurer}, \citenamefont {Stoyanov}, \citenamefont {Ward}, \citenamefont {Vaughan}, \citenamefont {Statz},\ and\ \citenamefont {Nelson}}]{nelson_review}%
  \BibitemOpen
  \bibfield  {author} {\bibinfo {author} {\bibfnamefont {T.}~\bibnamefont {Feurer}}, \bibinfo {author} {\bibfnamefont {N.~S.}\ \bibnamefont {Stoyanov}}, \bibinfo {author} {\bibfnamefont {D.~W.}\ \bibnamefont {Ward}}, \bibinfo {author} {\bibfnamefont {J.~C.}\ \bibnamefont {Vaughan}}, \bibinfo {author} {\bibfnamefont {E.~R.}\ \bibnamefont {Statz}}, \ and\ \bibinfo {author} {\bibfnamefont {K.~A.}\ \bibnamefont {Nelson}},\ }\href {\doibase 10.1146/annurev.matsci.37.052506.084327} {\bibfield  {journal} {\bibinfo  {journal} {Annual Review of Materials Research}\ }\textbf {\bibinfo {volume} {37}},\ \bibinfo {pages} {317} (\bibinfo {year} {2007}{\natexlab{a}})}\BibitemShut {NoStop}%
\bibitem [{\citenamefont {Feurer}\ \emph {et~al.}(2003)\citenamefont {Feurer}, \citenamefont {Vaughan},\ and\ \citenamefont {Nelson}}]{nelson_science03}%
  \BibitemOpen
  \bibfield  {author} {\bibinfo {author} {\bibfnamefont {T.}~\bibnamefont {Feurer}}, \bibinfo {author} {\bibfnamefont {J.~C.}\ \bibnamefont {Vaughan}}, \ and\ \bibinfo {author} {\bibfnamefont {K.~A.}\ \bibnamefont {Nelson}},\ }\href {\doibase 10.1126/science.1078726} {\bibfield  {journal} {\bibinfo  {journal} {Science}\ }\textbf {\bibinfo {volume} {299}},\ \bibinfo {pages} {374} (\bibinfo {year} {2003})}\BibitemShut {NoStop}%
\bibitem [{\citenamefont {Feurer}\ \emph {et~al.}(2007{\natexlab{b}})\citenamefont {Feurer}, \citenamefont {Stoyanov}, \citenamefont {Ward}, \citenamefont {Vaughan}, \citenamefont {Statz},\ and\ \citenamefont {Nelson}}]{nelson_review07}%
  \BibitemOpen
  \bibfield  {author} {\bibinfo {author} {\bibfnamefont {T.}~\bibnamefont {Feurer}}, \bibinfo {author} {\bibfnamefont {N.~S.}\ \bibnamefont {Stoyanov}}, \bibinfo {author} {\bibfnamefont {D.~W.}\ \bibnamefont {Ward}}, \bibinfo {author} {\bibfnamefont {J.~C.}\ \bibnamefont {Vaughan}}, \bibinfo {author} {\bibfnamefont {E.~R.}\ \bibnamefont {Statz}}, \ and\ \bibinfo {author} {\bibfnamefont {K.~A.}\ \bibnamefont {Nelson}},\ }\href {\doibase https://doi.org/10.1146/annurev.matsci.37.052506.084327} {\bibfield  {journal} {\bibinfo  {journal} {Annual Review of Materials Research}\ }\textbf {\bibinfo {volume} {37}},\ \bibinfo {pages} {317} (\bibinfo {year} {2007}{\natexlab{b}})}\BibitemShut {NoStop}%
\bibitem [{\citenamefont {Kojima}\ \emph {et~al.}(2002)\citenamefont {Kojima}, \citenamefont {Tsumura}, \citenamefont {Kitahara}, \citenamefont {Takeda},\ and\ \citenamefont {Nishizawa}}]{kojima_japplphys02}%
  \BibitemOpen
  \bibfield  {author} {\bibinfo {author} {\bibfnamefont {S.}~\bibnamefont {Kojima}}, \bibinfo {author} {\bibfnamefont {N.}~\bibnamefont {Tsumura}}, \bibinfo {author} {\bibfnamefont {H.}~\bibnamefont {Kitahara}}, \bibinfo {author} {\bibfnamefont {M.~W.}\ \bibnamefont {Takeda}}, \ and\ \bibinfo {author} {\bibfnamefont {S.}~\bibnamefont {Nishizawa}},\ }\href {\doibase 10.1143/JJAP.41.7033} {\bibfield  {journal} {\bibinfo  {journal} {Japanese Journal of Applied Physics}\ }\textbf {\bibinfo {volume} {41}},\ \bibinfo {pages} {7033} (\bibinfo {year} {2002})}\BibitemShut {NoStop}%
\bibitem [{\citenamefont {Cheung}\ and\ \citenamefont {Auston}(1985)}]{auston_prl85}%
  \BibitemOpen
  \bibfield  {author} {\bibinfo {author} {\bibfnamefont {K.~P.}\ \bibnamefont {Cheung}}\ and\ \bibinfo {author} {\bibfnamefont {D.~H.}\ \bibnamefont {Auston}},\ }\href {\doibase 10.1103/PhysRevLett.55.2152} {\bibfield  {journal} {\bibinfo  {journal} {Phys. Rev. Lett.}\ }\textbf {\bibinfo {volume} {55}},\ \bibinfo {pages} {2152} (\bibinfo {year} {1985})}\BibitemShut {NoStop}%
\bibitem [{\citenamefont {Dougherty}\ \emph {et~al.}(1992)\citenamefont {Dougherty}, \citenamefont {Wiederrecht},\ and\ \citenamefont {Nelson}}]{nelson_optica92}%
  \BibitemOpen
  \bibfield  {author} {\bibinfo {author} {\bibfnamefont {T.~P.}\ \bibnamefont {Dougherty}}, \bibinfo {author} {\bibfnamefont {G.~P.}\ \bibnamefont {Wiederrecht}}, \ and\ \bibinfo {author} {\bibfnamefont {K.~A.}\ \bibnamefont {Nelson}},\ }\href {\doibase 10.1364/JOSAB.9.002179} {\bibfield  {journal} {\bibinfo  {journal} {J. Opt. Soc. Am. B}\ }\textbf {\bibinfo {volume} {9}},\ \bibinfo {pages} {2179} (\bibinfo {year} {1992})}\BibitemShut {NoStop}%
\bibitem [{\citenamefont {Bakker}\ \emph {et~al.}(1992)\citenamefont {Bakker}, \citenamefont {Hunsche},\ and\ \citenamefont {Kurz}}]{kurz_prl92}%
  \BibitemOpen
  \bibfield  {author} {\bibinfo {author} {\bibfnamefont {H.~J.}\ \bibnamefont {Bakker}}, \bibinfo {author} {\bibfnamefont {S.}~\bibnamefont {Hunsche}}, \ and\ \bibinfo {author} {\bibfnamefont {H.}~\bibnamefont {Kurz}},\ }\href {\doibase 10.1103/PhysRevLett.69.2823} {\bibfield  {journal} {\bibinfo  {journal} {Phys. Rev. Lett.}\ }\textbf {\bibinfo {volume} {69}},\ \bibinfo {pages} {2823} (\bibinfo {year} {1992})}\BibitemShut {NoStop}%
\bibitem [{\citenamefont {Loulergue}\ and\ \citenamefont {Etchepare}(1995)}]{etchepare_prb95}%
  \BibitemOpen
  \bibfield  {author} {\bibinfo {author} {\bibfnamefont {J.~C.}\ \bibnamefont {Loulergue}}\ and\ \bibinfo {author} {\bibfnamefont {J.}~\bibnamefont {Etchepare}},\ }\href {\doibase 10.1103/PhysRevB.52.15160} {\bibfield  {journal} {\bibinfo  {journal} {Phys. Rev. B}\ }\textbf {\bibinfo {volume} {52}},\ \bibinfo {pages} {15160} (\bibinfo {year} {1995})}\BibitemShut {NoStop}%
\bibitem [{\citenamefont {Gautier}\ \emph {et~al.}(2000)\citenamefont {Gautier}, \citenamefont {Mérian},\ and\ \citenamefont {Etchepare}}]{etchepare_jpcm00}%
  \BibitemOpen
  \bibfield  {author} {\bibinfo {author} {\bibfnamefont {C.~A.}\ \bibnamefont {Gautier}}, \bibinfo {author} {\bibfnamefont {M.}~\bibnamefont {Mérian}}, \ and\ \bibinfo {author} {\bibfnamefont {J.}~\bibnamefont {Etchepare}},\ }\href {\doibase 10.1088/0953-8984/12/32/302} {\bibfield  {journal} {\bibinfo  {journal} {Journal of Physics: Condensed Matter}\ }\textbf {\bibinfo {volume} {12}},\ \bibinfo {pages} {7175} (\bibinfo {year} {2000})}\BibitemShut {NoStop}%
\bibitem [{\citenamefont {Knighton}\ \emph {et~al.}(2018)\citenamefont {Knighton}, \citenamefont {Dastrup}, \citenamefont {Johnson},\ and\ \citenamefont {Johnson}}]{johnson_prb18}%
  \BibitemOpen
  \bibfield  {author} {\bibinfo {author} {\bibfnamefont {B.~E.}\ \bibnamefont {Knighton}}, \bibinfo {author} {\bibfnamefont {B.~S.}\ \bibnamefont {Dastrup}}, \bibinfo {author} {\bibfnamefont {C.~L.}\ \bibnamefont {Johnson}}, \ and\ \bibinfo {author} {\bibfnamefont {J.~A.}\ \bibnamefont {Johnson}},\ }\href {\doibase 10.1103/PhysRevB.97.214307} {\bibfield  {journal} {\bibinfo  {journal} {Phys. Rev. B}\ }\textbf {\bibinfo {volume} {97}},\ \bibinfo {pages} {214307} (\bibinfo {year} {2018})}\BibitemShut {NoStop}%
\bibitem [{\citenamefont {Crimmins}\ \emph {et~al.}(2002)\citenamefont {Crimmins}, \citenamefont {Stoyanov},\ and\ \citenamefont {Nelson}}]{nelson_jcp02}%
  \BibitemOpen
  \bibfield  {author} {\bibinfo {author} {\bibfnamefont {T.~F.}\ \bibnamefont {Crimmins}}, \bibinfo {author} {\bibfnamefont {N.~S.}\ \bibnamefont {Stoyanov}}, \ and\ \bibinfo {author} {\bibfnamefont {K.~A.}\ \bibnamefont {Nelson}},\ }\href {\doibase 10.1063/1.1491948} {\bibfield  {journal} {\bibinfo  {journal} {The Journal of Chemical Physics}\ }\textbf {\bibinfo {volume} {117}},\ \bibinfo {pages} {2882} (\bibinfo {year} {2002})}\BibitemShut {NoStop}%
\bibitem [{\citenamefont {Hebling}(2002)}]{hebling_prb02}%
  \BibitemOpen
  \bibfield  {author} {\bibinfo {author} {\bibfnamefont {J.}~\bibnamefont {Hebling}},\ }\href {\doibase 10.1103/PhysRevB.65.092301} {\bibfield  {journal} {\bibinfo  {journal} {Phys. Rev. B}\ }\textbf {\bibinfo {volume} {65}},\ \bibinfo {pages} {092301} (\bibinfo {year} {2002})}\BibitemShut {NoStop}%
\bibitem [{\citenamefont {Wahlstrand}\ and\ \citenamefont {Merlin}(2003)}]{merlin_prb03}%
  \BibitemOpen
  \bibfield  {author} {\bibinfo {author} {\bibfnamefont {J.~K.}\ \bibnamefont {Wahlstrand}}\ and\ \bibinfo {author} {\bibfnamefont {R.}~\bibnamefont {Merlin}},\ }\href {\doibase 10.1103/PhysRevB.68.054301} {\bibfield  {journal} {\bibinfo  {journal} {Phys. Rev. B}\ }\textbf {\bibinfo {volume} {68}},\ \bibinfo {pages} {054301} (\bibinfo {year} {2003})}\BibitemShut {NoStop}%
\bibitem [{\citenamefont {Ikegaya}\ \emph {et~al.}(2015)\citenamefont {Ikegaya}, \citenamefont {Sakaibara}, \citenamefont {Minami}, \citenamefont {Katayama},\ and\ \citenamefont {Takeda}}]{takeda_apl15}%
  \BibitemOpen
  \bibfield  {author} {\bibinfo {author} {\bibfnamefont {Y.}~\bibnamefont {Ikegaya}}, \bibinfo {author} {\bibfnamefont {H.}~\bibnamefont {Sakaibara}}, \bibinfo {author} {\bibfnamefont {Y.}~\bibnamefont {Minami}}, \bibinfo {author} {\bibfnamefont {I.}~\bibnamefont {Katayama}}, \ and\ \bibinfo {author} {\bibfnamefont {J.}~\bibnamefont {Takeda}},\ }\href {\doibase 10.1063/1.4928480} {\bibfield  {journal} {\bibinfo  {journal} {Applied Physics Letters}\ }\textbf {\bibinfo {volume} {107}},\ \bibinfo {pages} {062901} (\bibinfo {year} {2015})}\BibitemShut {NoStop}%
\bibitem [{\citenamefont {Henry}\ and\ \citenamefont {Hopfield}(1965)}]{henry_prl65}%
  \BibitemOpen
  \bibfield  {author} {\bibinfo {author} {\bibfnamefont {C.~H.}\ \bibnamefont {Henry}}\ and\ \bibinfo {author} {\bibfnamefont {J.~J.}\ \bibnamefont {Hopfield}},\ }\href {\doibase 10.1103/PhysRevLett.15.964} {\bibfield  {journal} {\bibinfo  {journal} {Phys. Rev. Lett.}\ }\textbf {\bibinfo {volume} {15}},\ \bibinfo {pages} {964} (\bibinfo {year} {1965})}\BibitemShut {NoStop}%
\bibitem [{\citenamefont {Irmer}\ \emph {et~al.}(2013)\citenamefont {Irmer}, \citenamefont {R\"oder}, \citenamefont {Himcinschi},\ and\ \citenamefont {Kortus}}]{kortus_prb13}%
  \BibitemOpen
  \bibfield  {author} {\bibinfo {author} {\bibfnamefont {G.}~\bibnamefont {Irmer}}, \bibinfo {author} {\bibfnamefont {C.}~\bibnamefont {R\"oder}}, \bibinfo {author} {\bibfnamefont {C.}~\bibnamefont {Himcinschi}}, \ and\ \bibinfo {author} {\bibfnamefont {J.}~\bibnamefont {Kortus}},\ }\href {\doibase 10.1103/PhysRevB.88.104303} {\bibfield  {journal} {\bibinfo  {journal} {Phys. Rev. B}\ }\textbf {\bibinfo {volume} {88}},\ \bibinfo {pages} {104303} (\bibinfo {year} {2013})}\BibitemShut {NoStop}%
\bibitem [{\citenamefont {Barker}\ and\ \citenamefont {Loudon}(1972)}]{loudon_revmodphys72}%
  \BibitemOpen
  \bibfield  {author} {\bibinfo {author} {\bibfnamefont {A.~S.}\ \bibnamefont {Barker}}\ and\ \bibinfo {author} {\bibfnamefont {R.}~\bibnamefont {Loudon}},\ }\href {\doibase 10.1103/RevModPhys.44.18} {\bibfield  {journal} {\bibinfo  {journal} {Rev. Mod. Phys.}\ }\textbf {\bibinfo {volume} {44}},\ \bibinfo {pages} {18} (\bibinfo {year} {1972})}\BibitemShut {NoStop}%
\bibitem [{\citenamefont {Vicario}\ \emph {et~al.}(2020)\citenamefont {Vicario}, \citenamefont {Trisorio}, \citenamefont {Allenspach}, \citenamefont {Rüegg},\ and\ \citenamefont {Giorgianni}}]{giorgianni_narrow}%
  \BibitemOpen
  \bibfield  {author} {\bibinfo {author} {\bibfnamefont {C.}~\bibnamefont {Vicario}}, \bibinfo {author} {\bibfnamefont {A.}~\bibnamefont {Trisorio}}, \bibinfo {author} {\bibfnamefont {S.}~\bibnamefont {Allenspach}}, \bibinfo {author} {\bibfnamefont {C.}~\bibnamefont {Rüegg}}, \ and\ \bibinfo {author} {\bibfnamefont {F.}~\bibnamefont {Giorgianni}},\ }\href {\doibase 10.1063/5.0015612} {\bibfield  {journal} {\bibinfo  {journal} {Applied Physics Letters}\ }\textbf {\bibinfo {volume} {117}},\ \bibinfo {pages} {101101} (\bibinfo {year} {2020})}\BibitemShut {NoStop}%
\bibitem [{\citenamefont {Juv\'{e}}\ \emph {et~al.}(2018)\citenamefont {Juv\'{e}}, \citenamefont {Vaudel}, \citenamefont {Ollmann}, \citenamefont {Hebling}, \citenamefont {Temnov}, \citenamefont {Gusev},\ and\ \citenamefont {Pezeril}}]{pezeril_optica18}%
  \BibitemOpen
  \bibfield  {author} {\bibinfo {author} {\bibfnamefont {V.}~\bibnamefont {Juv\'{e}}}, \bibinfo {author} {\bibfnamefont {G.}~\bibnamefont {Vaudel}}, \bibinfo {author} {\bibfnamefont {Z.}~\bibnamefont {Ollmann}}, \bibinfo {author} {\bibfnamefont {J.}~\bibnamefont {Hebling}}, \bibinfo {author} {\bibfnamefont {V.}~\bibnamefont {Temnov}}, \bibinfo {author} {\bibfnamefont {V.}~\bibnamefont {Gusev}}, \ and\ \bibinfo {author} {\bibfnamefont {T.}~\bibnamefont {Pezeril}},\ }\href {\doibase 10.1364/OL.43.005905} {\bibfield  {journal} {\bibinfo  {journal} {Opt. Lett.}\ }\textbf {\bibinfo {volume} {43}},\ \bibinfo {pages} {5905} (\bibinfo {year} {2018})}\BibitemShut {NoStop}%
\bibitem [{\citenamefont {Huber}\ \emph {et~al.}(2015)\citenamefont {Huber}, \citenamefont {Ranke}, \citenamefont {Ferrer}, \citenamefont {Huber},\ and\ \citenamefont {Johnson}}]{johnson_apl15}%
  \BibitemOpen
  \bibfield  {author} {\bibinfo {author} {\bibfnamefont {T.}~\bibnamefont {Huber}}, \bibinfo {author} {\bibfnamefont {M.}~\bibnamefont {Ranke}}, \bibinfo {author} {\bibfnamefont {A.}~\bibnamefont {Ferrer}}, \bibinfo {author} {\bibfnamefont {L.}~\bibnamefont {Huber}}, \ and\ \bibinfo {author} {\bibfnamefont {S.~L.}\ \bibnamefont {Johnson}},\ }\href {\doibase 10.1063/1.4930021} {\bibfield  {journal} {\bibinfo  {journal} {Applied Physics Letters}\ }\textbf {\bibinfo {volume} {107}},\ \bibinfo {pages} {091107} (\bibinfo {year} {2015})}\BibitemShut {NoStop}%
\bibitem [{\citenamefont {Kojima}(2018)}]{kojima_photonics18}%
  \BibitemOpen
  \bibfield  {author} {\bibinfo {author} {\bibfnamefont {S.}~\bibnamefont {Kojima}},\ }\href {\doibase 10.3390/photonics5040055} {\bibfield  {journal} {\bibinfo  {journal} {Photonics}\ }\textbf {\bibinfo {volume} {5}} (\bibinfo {year} {2018}),\ 10.3390/photonics5040055}\BibitemShut {NoStop}%
\bibitem [{\citenamefont {Dastrup}\ \emph {et~al.}(2017)\citenamefont {Dastrup}, \citenamefont {Hall},\ and\ \citenamefont {Johnson}}]{johnson_apl17}%
  \BibitemOpen
  \bibfield  {author} {\bibinfo {author} {\bibfnamefont {B.~S.}\ \bibnamefont {Dastrup}}, \bibinfo {author} {\bibfnamefont {J.~R.}\ \bibnamefont {Hall}}, \ and\ \bibinfo {author} {\bibfnamefont {J.~A.}\ \bibnamefont {Johnson}},\ }\href {\doibase 10.1063/1.4980112} {\bibfield  {journal} {\bibinfo  {journal} {Applied Physics Letters}\ }\textbf {\bibinfo {volume} {110}},\ \bibinfo {pages} {162901} (\bibinfo {year} {2017})}\BibitemShut {NoStop}%
\bibitem [{\citenamefont {Johnson}\ \emph {et~al.}(2019)\citenamefont {Johnson}, \citenamefont {Knighton},\ and\ \citenamefont {Johnson}}]{johnson_prl19}%
  \BibitemOpen
  \bibfield  {author} {\bibinfo {author} {\bibfnamefont {C.~L.}\ \bibnamefont {Johnson}}, \bibinfo {author} {\bibfnamefont {B.~E.}\ \bibnamefont {Knighton}}, \ and\ \bibinfo {author} {\bibfnamefont {J.~A.}\ \bibnamefont {Johnson}},\ }\href {\doibase 10.1103/PhysRevLett.122.073901} {\bibfield  {journal} {\bibinfo  {journal} {Phys. Rev. Lett.}\ }\textbf {\bibinfo {volume} {122}},\ \bibinfo {pages} {073901} (\bibinfo {year} {2019})}\BibitemShut {NoStop}%
\bibitem [{\citenamefont {Knighton}\ \emph {et~al.}(2019)\citenamefont {Knighton}, \citenamefont {Tanner~Hardy}, \citenamefont {Johnson}, \citenamefont {Rawlings}, \citenamefont {Woolley}, \citenamefont {Calderon}, \citenamefont {Urrea},\ and\ \citenamefont {Johnson}}]{johnson_jap19}%
  \BibitemOpen
  \bibfield  {author} {\bibinfo {author} {\bibfnamefont {B.~E.}\ \bibnamefont {Knighton}}, \bibinfo {author} {\bibfnamefont {R.}~\bibnamefont {Tanner~Hardy}}, \bibinfo {author} {\bibfnamefont {C.~L.}\ \bibnamefont {Johnson}}, \bibinfo {author} {\bibfnamefont {L.~M.}\ \bibnamefont {Rawlings}}, \bibinfo {author} {\bibfnamefont {J.~T.}\ \bibnamefont {Woolley}}, \bibinfo {author} {\bibfnamefont {C.}~\bibnamefont {Calderon}}, \bibinfo {author} {\bibfnamefont {A.}~\bibnamefont {Urrea}}, \ and\ \bibinfo {author} {\bibfnamefont {J.~A.}\ \bibnamefont {Johnson}},\ }\href {\doibase 10.1063/1.5052638} {\bibfield  {journal} {\bibinfo  {journal} {Journal of Applied Physics}\ }\textbf {\bibinfo {volume} {125}},\ \bibinfo {pages} {144101} (\bibinfo {year} {2019})}\BibitemShut {NoStop}%
\bibitem [{\citenamefont {Lin}\ \emph {et~al.}(2022)\citenamefont {Lin}, \citenamefont {Mead},\ and\ \citenamefont {Blake}}]{blake_prl22}%
  \BibitemOpen
  \bibfield  {author} {\bibinfo {author} {\bibfnamefont {H.-W.}\ \bibnamefont {Lin}}, \bibinfo {author} {\bibfnamefont {G.}~\bibnamefont {Mead}}, \ and\ \bibinfo {author} {\bibfnamefont {G.~A.}\ \bibnamefont {Blake}},\ }\href {\doibase 10.1103/PhysRevLett.129.207401} {\bibfield  {journal} {\bibinfo  {journal} {Phys. Rev. Lett.}\ }\textbf {\bibinfo {volume} {129}},\ \bibinfo {pages} {207401} (\bibinfo {year} {2022})}\BibitemShut {NoStop}%
\bibitem [{\citenamefont {Fu}\ and\ \citenamefont {Yamaguchi}(2016)}]{yamaguchi_scirep16}%
  \BibitemOpen
  \bibfield  {author} {\bibinfo {author} {\bibfnamefont {Z.}~\bibnamefont {Fu}}\ and\ \bibinfo {author} {\bibfnamefont {M.}~\bibnamefont {Yamaguchi}},\ }\href {\doibase 10.1038/srep38264} {\bibfield  {journal} {\bibinfo  {journal} {Scientific Reports}\ }\textbf {\bibinfo {volume} {6}},\ \bibinfo {pages} {38264} (\bibinfo {year} {2016})}\BibitemShut {NoStop}%
\bibitem [{\citenamefont {Faust}\ and\ \citenamefont {Henry}(1966)}]{henry_prl66}%
  \BibitemOpen
  \bibfield  {author} {\bibinfo {author} {\bibfnamefont {W.~L.}\ \bibnamefont {Faust}}\ and\ \bibinfo {author} {\bibfnamefont {C.~H.}\ \bibnamefont {Henry}},\ }\href {\doibase 10.1103/PhysRevLett.17.1265} {\bibfield  {journal} {\bibinfo  {journal} {Phys. Rev. Lett.}\ }\textbf {\bibinfo {volume} {17}},\ \bibinfo {pages} {1265} (\bibinfo {year} {1966})}\BibitemShut {NoStop}%
\bibitem [{\citenamefont {Hwang}\ and\ \citenamefont {Solin}(1973)}]{hwang_ssc73}%
  \BibitemOpen
  \bibfield  {author} {\bibinfo {author} {\bibfnamefont {D.}~\bibnamefont {Hwang}}\ and\ \bibinfo {author} {\bibfnamefont {S.}~\bibnamefont {Solin}},\ }\href {\doibase https://doi.org/10.1016/0038-1098(73)90414-6} {\bibfield  {journal} {\bibinfo  {journal} {Solid State Communications}\ }\textbf {\bibinfo {volume} {13}},\ \bibinfo {pages} {983} (\bibinfo {year} {1973})}\BibitemShut {NoStop}%
\bibitem [{\citenamefont {Hwang}\ and\ \citenamefont {Solin}(1974)}]{hwang_prb74}%
  \BibitemOpen
  \bibfield  {author} {\bibinfo {author} {\bibfnamefont {D.~M.}\ \bibnamefont {Hwang}}\ and\ \bibinfo {author} {\bibfnamefont {S.~A.}\ \bibnamefont {Solin}},\ }\href {\doibase 10.1103/PhysRevB.9.1884} {\bibfield  {journal} {\bibinfo  {journal} {Phys. Rev. B}\ }\textbf {\bibinfo {volume} {9}},\ \bibinfo {pages} {1884} (\bibinfo {year} {1974})}\BibitemShut {NoStop}%
\bibitem [{\citenamefont {Wagner}\ and\ \citenamefont {Claus}(1981)}]{claus_pssa81}%
  \BibitemOpen
  \bibfield  {author} {\bibinfo {author} {\bibfnamefont {W.~D.}\ \bibnamefont {Wagner}}\ and\ \bibinfo {author} {\bibfnamefont {R.}~\bibnamefont {Claus}},\ }\href {\doibase https://doi.org/10.1002/pssa.2210640229} {\bibfield  {journal} {\bibinfo  {journal} {physica status solidi (a)}\ }\textbf {\bibinfo {volume} {64}},\ \bibinfo {pages} {647} (\bibinfo {year} {1981})}\BibitemShut {NoStop}%
\bibitem [{\citenamefont {Mayer}\ and\ \citenamefont {Keilmann}(1986)}]{keilmann_prb86}%
  \BibitemOpen
  \bibfield  {author} {\bibinfo {author} {\bibfnamefont {A.}~\bibnamefont {Mayer}}\ and\ \bibinfo {author} {\bibfnamefont {F.}~\bibnamefont {Keilmann}},\ }\href {\doibase 10.1103/PhysRevB.33.6954} {\bibfield  {journal} {\bibinfo  {journal} {Phys. Rev. B}\ }\textbf {\bibinfo {volume} {33}},\ \bibinfo {pages} {6954} (\bibinfo {year} {1986})}\BibitemShut {NoStop}%
\bibitem [{\citenamefont {Dekorsy}\ \emph {et~al.}(2003)\citenamefont {Dekorsy}, \citenamefont {Yakovlev}, \citenamefont {Seidel}, \citenamefont {Helm},\ and\ \citenamefont {Keilmann}}]{keilmann_prl03}%
  \BibitemOpen
  \bibfield  {author} {\bibinfo {author} {\bibfnamefont {T.}~\bibnamefont {Dekorsy}}, \bibinfo {author} {\bibfnamefont {V.~A.}\ \bibnamefont {Yakovlev}}, \bibinfo {author} {\bibfnamefont {W.}~\bibnamefont {Seidel}}, \bibinfo {author} {\bibfnamefont {M.}~\bibnamefont {Helm}}, \ and\ \bibinfo {author} {\bibfnamefont {F.}~\bibnamefont {Keilmann}},\ }\href {\doibase 10.1103/PhysRevLett.90.055508} {\bibfield  {journal} {\bibinfo  {journal} {Phys. Rev. Lett.}\ }\textbf {\bibinfo {volume} {90}},\ \bibinfo {pages} {055508} (\bibinfo {year} {2003})}\BibitemShut {NoStop}%
\bibitem [{\citenamefont {Wiederrecht}\ \emph {et~al.}(1995)\citenamefont {Wiederrecht}, \citenamefont {Dougherty}, \citenamefont {Dhar}, \citenamefont {Nelson}, \citenamefont {Leaird},\ and\ \citenamefont {Weiner}}]{weiner_prb95}%
  \BibitemOpen
  \bibfield  {author} {\bibinfo {author} {\bibfnamefont {G.~P.}\ \bibnamefont {Wiederrecht}}, \bibinfo {author} {\bibfnamefont {T.~P.}\ \bibnamefont {Dougherty}}, \bibinfo {author} {\bibfnamefont {L.}~\bibnamefont {Dhar}}, \bibinfo {author} {\bibfnamefont {K.~A.}\ \bibnamefont {Nelson}}, \bibinfo {author} {\bibfnamefont {D.~E.}\ \bibnamefont {Leaird}}, \ and\ \bibinfo {author} {\bibfnamefont {A.~M.}\ \bibnamefont {Weiner}},\ }\href {\doibase 10.1103/PhysRevB.51.916} {\bibfield  {journal} {\bibinfo  {journal} {Phys. Rev. B}\ }\textbf {\bibinfo {volume} {51}},\ \bibinfo {pages} {916} (\bibinfo {year} {1995})}\BibitemShut {NoStop}%
\bibitem [{\citenamefont {Bakker}\ \emph {et~al.}(1994)\citenamefont {Bakker}, \citenamefont {Hunsche},\ and\ \citenamefont {Kurz}}]{kurz_prb94}%
  \BibitemOpen
  \bibfield  {author} {\bibinfo {author} {\bibfnamefont {H.~J.}\ \bibnamefont {Bakker}}, \bibinfo {author} {\bibfnamefont {S.}~\bibnamefont {Hunsche}}, \ and\ \bibinfo {author} {\bibfnamefont {H.}~\bibnamefont {Kurz}},\ }\href {\doibase 10.1103/PhysRevB.50.914} {\bibfield  {journal} {\bibinfo  {journal} {Phys. Rev. B}\ }\textbf {\bibinfo {volume} {50}},\ \bibinfo {pages} {914} (\bibinfo {year} {1994})}\BibitemShut {NoStop}%
\bibitem [{\citenamefont {Bakker}\ \emph {et~al.}(1998)\citenamefont {Bakker}, \citenamefont {Hunsche},\ and\ \citenamefont {Kurz}}]{kurz_RevModPhys98}%
  \BibitemOpen
  \bibfield  {author} {\bibinfo {author} {\bibfnamefont {H.~J.}\ \bibnamefont {Bakker}}, \bibinfo {author} {\bibfnamefont {S.}~\bibnamefont {Hunsche}}, \ and\ \bibinfo {author} {\bibfnamefont {H.}~\bibnamefont {Kurz}},\ }\href {\doibase 10.1103/RevModPhys.70.523} {\bibfield  {journal} {\bibinfo  {journal} {Rev. Mod. Phys.}\ }\textbf {\bibinfo {volume} {70}},\ \bibinfo {pages} {523} (\bibinfo {year} {1998})}\BibitemShut {NoStop}%
\bibitem [{\citenamefont {Cavalleri}\ \emph {et~al.}(2006)\citenamefont {Cavalleri}, \citenamefont {Wall}, \citenamefont {Simpson}, \citenamefont {Statz}, \citenamefont {Ward}, \citenamefont {Nelson}, \citenamefont {Rini},\ and\ \citenamefont {Schoenlein}}]{cavalleri_xray_nature06}%
  \BibitemOpen
  \bibfield  {author} {\bibinfo {author} {\bibfnamefont {A.}~\bibnamefont {Cavalleri}}, \bibinfo {author} {\bibfnamefont {S.}~\bibnamefont {Wall}}, \bibinfo {author} {\bibfnamefont {C.}~\bibnamefont {Simpson}}, \bibinfo {author} {\bibfnamefont {E.}~\bibnamefont {Statz}}, \bibinfo {author} {\bibfnamefont {D.~W.}\ \bibnamefont {Ward}}, \bibinfo {author} {\bibfnamefont {K.~A.}\ \bibnamefont {Nelson}}, \bibinfo {author} {\bibfnamefont {M.}~\bibnamefont {Rini}}, \ and\ \bibinfo {author} {\bibfnamefont {R.~W.}\ \bibnamefont {Schoenlein}},\ }\href {\doibase 10.1038/nature05041} {\bibfield  {journal} {\bibinfo  {journal} {Nature}\ }\textbf {\bibinfo {volume} {442}},\ \bibinfo {pages} {664} (\bibinfo {year} {2006})}\BibitemShut {NoStop}%
\bibitem [{\citenamefont {Mankowsky}\ \emph {et~al.}(2017)\citenamefont {Mankowsky}, \citenamefont {von Hoegen}, \citenamefont {F\"orst},\ and\ \citenamefont {Cavalleri}}]{cavalleri_prl17}%
  \BibitemOpen
  \bibfield  {author} {\bibinfo {author} {\bibfnamefont {R.}~\bibnamefont {Mankowsky}}, \bibinfo {author} {\bibfnamefont {A.}~\bibnamefont {von Hoegen}}, \bibinfo {author} {\bibfnamefont {M.}~\bibnamefont {F\"orst}}, \ and\ \bibinfo {author} {\bibfnamefont {A.}~\bibnamefont {Cavalleri}},\ }\href {\doibase 10.1103/PhysRevLett.118.197601} {\bibfield  {journal} {\bibinfo  {journal} {Phys. Rev. Lett.}\ }\textbf {\bibinfo {volume} {118}},\ \bibinfo {pages} {197601} (\bibinfo {year} {2017})}\BibitemShut {NoStop}%
\bibitem [{\citenamefont {Henstridge}\ \emph {et~al.}(2022)\citenamefont {Henstridge}, \citenamefont {Först}, \citenamefont {Rowe}, \citenamefont {Fechner},\ and\ \citenamefont {Cavalleri}}]{cavalleri_natphys22}%
  \BibitemOpen
  \bibfield  {author} {\bibinfo {author} {\bibfnamefont {M.}~\bibnamefont {Henstridge}}, \bibinfo {author} {\bibfnamefont {M.}~\bibnamefont {Först}}, \bibinfo {author} {\bibfnamefont {E.}~\bibnamefont {Rowe}}, \bibinfo {author} {\bibfnamefont {M.}~\bibnamefont {Fechner}}, \ and\ \bibinfo {author} {\bibfnamefont {A.}~\bibnamefont {Cavalleri}},\ }\href {https://doi.org/10.1038/s41567-022-01512-3} {\bibfield  {journal} {\bibinfo  {journal} {Nature Physics}\ }\textbf {\bibinfo {volume} {18}},\ \bibinfo {pages} {461} (\bibinfo {year} {2022})}\BibitemShut {NoStop}%
\bibitem [{\citenamefont {Taherian}\ \emph {et~al.}(2024)\citenamefont {Taherian}, \citenamefont {Först}, \citenamefont {Liu}, \citenamefont {Fechner}, \citenamefont {Pavicevic}, \citenamefont {von Hoegen}, \citenamefont {Rowe}, \citenamefont {Liu}, \citenamefont {Nakata}, \citenamefont {Keimer}, \citenamefont {Demler}, \citenamefont {Michael},\ and\ \citenamefont {Cavalleri}}]{cavalleri_cm24}%
  \BibitemOpen
  \bibfield  {author} {\bibinfo {author} {\bibfnamefont {N.}~\bibnamefont {Taherian}}, \bibinfo {author} {\bibfnamefont {M.}~\bibnamefont {Först}}, \bibinfo {author} {\bibfnamefont {A.}~\bibnamefont {Liu}}, \bibinfo {author} {\bibfnamefont {M.}~\bibnamefont {Fechner}}, \bibinfo {author} {\bibfnamefont {D.}~\bibnamefont {Pavicevic}}, \bibinfo {author} {\bibfnamefont {A.}~\bibnamefont {von Hoegen}}, \bibinfo {author} {\bibfnamefont {E.}~\bibnamefont {Rowe}}, \bibinfo {author} {\bibfnamefont {Y.}~\bibnamefont {Liu}}, \bibinfo {author} {\bibfnamefont {S.}~\bibnamefont {Nakata}}, \bibinfo {author} {\bibfnamefont {B.}~\bibnamefont {Keimer}}, \bibinfo {author} {\bibfnamefont {E.}~\bibnamefont {Demler}}, \bibinfo {author} {\bibfnamefont {M.~H.}\ \bibnamefont {Michael}}, \ and\ \bibinfo {author} {\bibfnamefont {A.}~\bibnamefont {Cavalleri}},\ }\href@noop {} {\enquote {\bibinfo {title} {Squeezed josephson plasmons in driven $\text{YBa}_2\text{Cu}_3\text{O}_{6+x}$},}\ } (\bibinfo {year} {2024}),\ \Eprint
  {http://arxiv.org/abs/2401.01115} {arXiv:2401.01115 [cond-mat.supr-con]} \BibitemShut {NoStop}%
\bibitem [{\citenamefont {Luo}\ \emph {et~al.}(2024)\citenamefont {Luo}, \citenamefont {Ilyas}, \citenamefont {Hoegen}, \citenamefont {Lee}, \citenamefont {Park}, \citenamefont {Park},\ and\ \citenamefont {Gedik}}]{gedik_natcomm24}%
  \BibitemOpen
  \bibfield  {author} {\bibinfo {author} {\bibfnamefont {T.}~\bibnamefont {Luo}}, \bibinfo {author} {\bibfnamefont {B.}~\bibnamefont {Ilyas}}, \bibinfo {author} {\bibfnamefont {A.~v.}\ \bibnamefont {Hoegen}}, \bibinfo {author} {\bibfnamefont {Y.}~\bibnamefont {Lee}}, \bibinfo {author} {\bibfnamefont {J.}~\bibnamefont {Park}}, \bibinfo {author} {\bibfnamefont {J.-G.}\ \bibnamefont {Park}}, \ and\ \bibinfo {author} {\bibfnamefont {N.}~\bibnamefont {Gedik}},\ }\href {https://doi.org/10.1038/s41467-024-46515-1} {\bibfield  {journal} {\bibinfo  {journal} {Nature Communications}\ }\textbf {\bibinfo {volume} {15}},\ \bibinfo {pages} {2276} (\bibinfo {year} {2024})}\BibitemShut {NoStop}%
\bibitem [{\citenamefont {Nagaosa}\ and\ \citenamefont {Heusler}(1999)}]{nagaosa}%
  \BibitemOpen
  \bibfield  {author} {\bibinfo {author} {\bibfnamefont {N.}~\bibnamefont {Nagaosa}}\ and\ \bibinfo {author} {\bibfnamefont {S.}~\bibnamefont {Heusler}},\ }\href {https://books.google.it/books?id=C9uAXYIlFhMC} {\emph {\bibinfo {title} {Quantum Field Theory in Condensed Matter Physics}}},\ Texts and monographs in physics\ (\bibinfo  {publisher} {Springer, New York, NY},\ \bibinfo {year} {1999})\BibitemShut {NoStop}%
\bibitem [{\citenamefont {Mahan}(2000)}]{mahan}%
  \BibitemOpen
  \bibfield  {author} {\bibinfo {author} {\bibfnamefont {G.~D.}\ \bibnamefont {Mahan}},\ }\href@noop {} {\emph {\bibinfo {title} {Many Particle Physics, Third Edition}}}\ (\bibinfo  {publisher} {Plenum},\ \bibinfo {address} {New York},\ \bibinfo {year} {2000})\BibitemShut {NoStop}%
\bibitem [{\citenamefont {Udina}\ \emph {et~al.}(2019)\citenamefont {Udina}, \citenamefont {Cea},\ and\ \citenamefont {Benfatto}}]{udina_prb19}%
  \BibitemOpen
  \bibfield  {author} {\bibinfo {author} {\bibfnamefont {M.}~\bibnamefont {Udina}}, \bibinfo {author} {\bibfnamefont {T.}~\bibnamefont {Cea}}, \ and\ \bibinfo {author} {\bibfnamefont {L.}~\bibnamefont {Benfatto}},\ }\href {\doibase 10.1103/PhysRevB.100.165131} {\bibfield  {journal} {\bibinfo  {journal} {Phys. Rev. B}\ }\textbf {\bibinfo {volume} {100}},\ \bibinfo {pages} {165131} (\bibinfo {year} {2019})}\BibitemShut {NoStop}%
\bibitem [{sup()}]{suppl}%
  \BibitemOpen
  \href@noop {} {}\bibinfo {howpublished} {See Supplementary Material for the derivation of the dressed-phonon propagator and of the nonlinear current in the TWM and FWM case; considerations on the tensorial structure of the TWM kernel; derivation of the nonlinear signal through propagation effects in the pump-probe protocol and the effect of the sample thickness on the response.}\BibitemShut {Stop}%
\bibitem [{\citenamefont {Sivarajah}\ \emph {et~al.}(2019)\citenamefont {Sivarajah}, \citenamefont {Steinbacher}, \citenamefont {Dastrup}, \citenamefont {Lu}, \citenamefont {Xiang}, \citenamefont {Ren}, \citenamefont {Kamba}, \citenamefont {Cao},\ and\ \citenamefont {Nelson}}]{nelson_magn_pol}%
  \BibitemOpen
  \bibfield  {author} {\bibinfo {author} {\bibfnamefont {P.}~\bibnamefont {Sivarajah}}, \bibinfo {author} {\bibfnamefont {A.}~\bibnamefont {Steinbacher}}, \bibinfo {author} {\bibfnamefont {B.}~\bibnamefont {Dastrup}}, \bibinfo {author} {\bibfnamefont {J.}~\bibnamefont {Lu}}, \bibinfo {author} {\bibfnamefont {M.}~\bibnamefont {Xiang}}, \bibinfo {author} {\bibfnamefont {W.}~\bibnamefont {Ren}}, \bibinfo {author} {\bibfnamefont {S.}~\bibnamefont {Kamba}}, \bibinfo {author} {\bibfnamefont {S.}~\bibnamefont {Cao}}, \ and\ \bibinfo {author} {\bibfnamefont {K.~A.}\ \bibnamefont {Nelson}},\ }\href {\doibase 10.1063/1.5083849} {\bibfield  {journal} {\bibinfo  {journal} {Journal of Applied Physics}\ }\textbf {\bibinfo {volume} {125}},\ \bibinfo {pages} {213103} (\bibinfo {year} {2019})}\BibitemShut {NoStop}%
\bibitem [{\citenamefont {F{\"o}rst}\ \emph {et~al.}(2011)\citenamefont {F{\"o}rst}, \citenamefont {Manzoni}, \citenamefont {Kaiser}, \citenamefont {Tomioka}, \citenamefont {Tokura}, \citenamefont {Merlin},\ and\ \citenamefont {Cavalleri}}]{forst2011}%
  \BibitemOpen
  \bibfield  {author} {\bibinfo {author} {\bibfnamefont {M.}~\bibnamefont {F{\"o}rst}}, \bibinfo {author} {\bibfnamefont {C.}~\bibnamefont {Manzoni}}, \bibinfo {author} {\bibfnamefont {S.}~\bibnamefont {Kaiser}}, \bibinfo {author} {\bibfnamefont {Y.}~\bibnamefont {Tomioka}}, \bibinfo {author} {\bibfnamefont {Y.}~\bibnamefont {Tokura}}, \bibinfo {author} {\bibfnamefont {R.}~\bibnamefont {Merlin}}, \ and\ \bibinfo {author} {\bibfnamefont {A.}~\bibnamefont {Cavalleri}},\ }\href@noop {} {\bibfield  {journal} {\bibinfo  {journal} {Nature Physics}\ }\textbf {\bibinfo {volume} {7}},\ \bibinfo {pages} {854} (\bibinfo {year} {2011})}\BibitemShut {NoStop}%
\bibitem [{\citenamefont {Basini}\ \emph {et~al.}(2024)\citenamefont {Basini}, \citenamefont {Udina}, \citenamefont {Pancaldi}, \citenamefont {Unikandanunni}, \citenamefont {Bonetti},\ and\ \citenamefont {Benfatto}}]{IKE_24}%
  \BibitemOpen
  \bibfield  {author} {\bibinfo {author} {\bibfnamefont {M.}~\bibnamefont {Basini}}, \bibinfo {author} {\bibfnamefont {M.}~\bibnamefont {Udina}}, \bibinfo {author} {\bibfnamefont {M.}~\bibnamefont {Pancaldi}}, \bibinfo {author} {\bibfnamefont {V.}~\bibnamefont {Unikandanunni}}, \bibinfo {author} {\bibfnamefont {S.}~\bibnamefont {Bonetti}}, \ and\ \bibinfo {author} {\bibfnamefont {L.}~\bibnamefont {Benfatto}},\ }\href {\doibase 10.1103/PhysRevB.109.024309} {\bibfield  {journal} {\bibinfo  {journal} {Phys. Rev. B}\ }\textbf {\bibinfo {volume} {109}},\ \bibinfo {pages} {024309} (\bibinfo {year} {2024})}\BibitemShut {NoStop}%
\bibitem [{\citenamefont {DelPo}\ \emph {et~al.}(2020)\citenamefont {DelPo}, \citenamefont {Kudisch}, \citenamefont {Park}, \citenamefont {Khan}, \citenamefont {Fassioli}, \citenamefont {Fausti}, \citenamefont {Rand},\ and\ \citenamefont {Scholes}}]{delpo20}%
  \BibitemOpen
  \bibfield  {author} {\bibinfo {author} {\bibfnamefont {C.~A.}\ \bibnamefont {DelPo}}, \bibinfo {author} {\bibfnamefont {B.}~\bibnamefont {Kudisch}}, \bibinfo {author} {\bibfnamefont {K.~H.}\ \bibnamefont {Park}}, \bibinfo {author} {\bibfnamefont {S.-U.-Z.}\ \bibnamefont {Khan}}, \bibinfo {author} {\bibfnamefont {F.}~\bibnamefont {Fassioli}}, \bibinfo {author} {\bibfnamefont {D.}~\bibnamefont {Fausti}}, \bibinfo {author} {\bibfnamefont {B.~P.}\ \bibnamefont {Rand}}, \ and\ \bibinfo {author} {\bibfnamefont {G.~D.}\ \bibnamefont {Scholes}},\ }\href {\doibase 10.1021/acs.jpclett.0c00247} {\bibfield  {journal} {\bibinfo  {journal} {The Journal of Physical Chemistry Letters}\ }\textbf {\bibinfo {volume} {11}},\ \bibinfo {pages} {2667} (\bibinfo {year} {2020})},\ \bibinfo {note} {pMID: 32186878},\ \Eprint {http://arxiv.org/abs/https://doi.org/10.1021/acs.jpclett.0c00247} {https://doi.org/10.1021/acs.jpclett.0c00247} \BibitemShut {NoStop}%
\bibitem [{\citenamefont {Fassioli}\ \emph {et~al.}(2021)\citenamefont {Fassioli}, \citenamefont {Park}, \citenamefont {Bard},\ and\ \citenamefont {Scholes}}]{fassioli_jpcl21}%
  \BibitemOpen
  \bibfield  {author} {\bibinfo {author} {\bibfnamefont {F.}~\bibnamefont {Fassioli}}, \bibinfo {author} {\bibfnamefont {K.~H.}\ \bibnamefont {Park}}, \bibinfo {author} {\bibfnamefont {S.~E.}\ \bibnamefont {Bard}}, \ and\ \bibinfo {author} {\bibfnamefont {G.~D.}\ \bibnamefont {Scholes}},\ }\href {\doibase 10.1021/acs.jpclett.1c03183} {\bibfield  {journal} {\bibinfo  {journal} {The Journal of Physical Chemistry Letters}\ }\textbf {\bibinfo {volume} {12}},\ \bibinfo {pages} {11444} (\bibinfo {year} {2021})},\ \bibinfo {note} {pMID: 34792371}\BibitemShut {NoStop}%
\end{thebibliography}%

\end{document}